\documentclass[11pt,twoside]{article}
\usepackage{macroaaa-eng}
\usepackage{psfig}
\usepackage{graphicx}
\usepackage[T1]{fontenc} 

\usepackage{latexsym}
\usepackage{verbatim}

\begin{document}

\vskip 1.0cm
\markboth{Gustavo E. Romero}{Black holes}
\pagestyle{myheadings}

\vspace*{0.5cm}
\title{Introduction to black holes}

\author{Gustavo E. Romero}
\affil{Instituto Argentino de Radioastronom\'\i a, C.C. 5, Villa
Elisa (1894), and Facultad de Cs. Astron\'omicas y
Geof\'{\i}sicas, UNLP, Paseo del Bosque S/N, (1900) La Plata,
Argentina, romero@fcaglp.unlp.edu.ar}

\begin{abstract} 
Black holes are perhaps the most strange and fascinating objects known to exist in the universe. Our understanding of space and time is pushed to its limits by the extreme conditions found in these objects. They can be used as natural laboratories to test the behavior of matter in very strong gravitational fields. Black holes seem to play a key role in the universe, powering a wide variety of phenomena, from X-ray binaries to active galactic nuclei. In this article we will review the basics of black hole physics.        

\end{abstract}

\section{Introduction}

Strictly speaking, black holes do not exist. Actually, holes, of any kind, do not exist. You can talk about holes of course. For instance you can say: ``there is a hole in the wall''. You can give many details of the hole: it is big, it is round shaped, light comes in through it, even perhaps the hole could be such that you can go through to the outside. But I am sure that you do not think that there is a thing made out of nothingness in the wall. No. To talk about the hole is actually an indirect way of talking about the wall. What really exists is the wall. The wall is made out of bricks, atoms, protons and leptons, whatever. To say that there is a hole in the wall is just to say that the wall has a certain topology, a topology such that not every closed curve on the surface of the wall can be contracted to a single point. 
The hole is not a thing. The hole is a property of the wall. 

Let us come back to black holes. What are we talking about when we talk about black holes?. {\sl Spacetime}. OK, but what is spacetime?. {\sl Spacetime is the set of all events of all things}. Everything that has happened, everything that happens, everything that will happen, is just an element, a ``point'', of spacetime. Spacetime is not a thing, it is just the way we represent the relations among all things. 

As with every set, we can equip spacetime with some mathematical structure, in order to describe it. We shall adopt the following mathematical structure for spacetime:\\

{\sl Spacetime is represented by a differentiable, 4-dimensional, real manifold.}\\ 

A real 4-D manifold is a set that can be covered completely by subsets whose elements are in a one-to-one correspondence with subsets of $\Re^{4}$. Each element of the manifold represents an event. We adopt 4 dimensions because it seems enough to give 4 real numbers to localize an event. For instance, a lightning has beaten the top of the building, located in the corner of streets 115 and 46, at 25 m above the see level, La Plata city, at 4.35 am, local time, March 2nd, 2008 (this is my home, by the way). We see now why we choose a manifold to represent spacetime: we can always provide a set of 4 real numbers for every event, and this can be done independently of the geometric structure of the manifold. If there is more than a single characterization of an event, we can always find a transformation law between the different coordinate systems. This is a basic property of manifolds. 

Now, if we want to calculate distances between two events is spacetime, we need more structure on our manifold: we need a geometric structure. We can do this introducing a metric tensor that tell us how to calculate distances. For instance, consider an Euclidean metric tensor $\delta_{\mu\nu}$ (indices run from 0 to 3):

$$	\delta_{\mu\nu} = \left(\matrix{1 &0&0&0\cr 0&1&0&0\cr 
  0&0&1&0 \cr 0&0&0&1\cr}\right)$$

Then, adopting the Einstein convention of sum, we have that the distance $ds$ between two arbitrarily close events is: 
$$ds^{2}=\delta_{\mu\nu} dx^{\mu} dx^{\nu}=(dx^{0})^{2}+(dx^{1})^{2}+(dx^{3})^{2}+(dx^{3})^{2}.$$ Restricted to 3 coordinates, this is the way distances have been calculated since Pythagoras. However, the world seems to be a little more complicated. After the introduction of the Special Theory of Relativity by Einstein (1905), the German mathematician Hermann Minkowski introduced the following pseudo-Euclidean metric which is consistent with Einstein's theory (Minkowski 1907):

\begin{equation}
	ds^{2}=\eta_{\mu\nu}dx^{\mu} dx^{\nu}=(dx^{0})^{2}-(dx^{1})^{2}-(dx^{3})^{2}-(dx^{3})^{2}. \label{eta}
\end{equation}

Notice that the Minkowski metric tensor $\eta_{\mu\nu}$ has rank 2 and trace $-2$. We call the coordinates with the same sign {\sl spatial} (adopting the convention $x^{1}=x$,  $x^{2}=y$, and $x^{3}=z$) and the coordinate $x^{0}=ct$ is called {\sl temporal coordinate}. The constant $c$ is introduced to make uniform the units. There 
is an important fact respect to Eq. (\ref{eta}): contrary to what was thought by Kant and others, it is not a necessary statement. Things might have been different. We can easily imagine possible worlds with other metrics.  This means that the metric tensor has empirical information about the real universe. 

\begin{figure}  
\includegraphics[trim=0 10 0 0,clip,width=14cm]{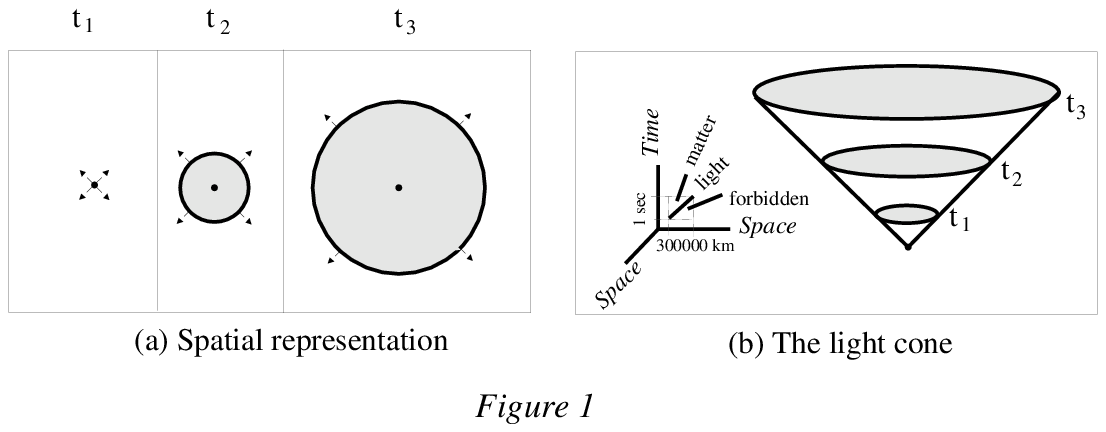}

\caption{Light cone. From J-P. Luminet (1998).}\label{cone}
\end{figure}

Once we have introduced a metric tensor we can separate spacetime at each point in three regions according to $ds^{2}<0$ (space-like region), $ds^{2}=0$ (light-like or null region), and $ds^{2}>0$ (time-like region). Particles that go through the origin can only reach time-like regions. The null surface $ds^{2}=0$ can be inhabited only by particles moving at the speed of light, like photons. Points in the space-like region cannot been reached by material objects from the origin of the {\sl light cone} that can be formed at any spacetime point. Notice that the introduction of the metric allows to define the future and the past {of a given event}. Let us consider the unitary vector $T^{\nu}=(1,\;0,\;0,\;0)$, then a vector $x^{\mu}$ points to the future if $\eta_{\mu\nu}   x^{\mu} T^{\nu}>0$. In the similar way, the vector points toward the past if $\eta_{\mu\nu}   x^{\mu} T^{\nu}<0$. A light cone is shown in Figure \ref{cone}.

We define the {\sl proper time} ($\tau$) of a physical system as the time of a co-moving system, i.e. $dx=dy=dz=0$, and hence:
\begin{equation}
	d\tau^{2}=\frac{1}{c^{2}}ds^{2}.
\end{equation}
Since the interval is an invariant (i.e. it has the same value in all coordinate systems), it is easy to show that:
\begin{equation}
	d\tau=\frac{dt}{\gamma},
\end{equation}
where
\begin{equation}
	\gamma=\frac{1}{\sqrt{1-\left(\frac{v}{c}\right)^2}}
\end{equation}
is the Lorentz factor of the system. 

A basic characteristic of Minskowski spacetime is that it is ``flat'': all light cones point in the same direction, i.e. the local direction of the future does not depend on the coefficients of the metric since these are constants. More general spacetimes are possible. If we want to describe gravity in the framework of spacetime, we have to introduce a Riemannian spacetime, whose metric can be flexible, i.e. a function of the material properties (mass-energy and momentum) of the physical systems that produce the events of spacetime. 

In order to introduce gravitation in a general spacetime, we define a metric tensor $g_{\mu\nu}$, such as its components can be related to those of a locally Minkowski spacetime defined by $ds^{2}=\eta_{\alpha\beta}d\xi^{\alpha} d\xi^{\beta}$ through a general transformation:
\begin{equation}
	ds^{2}=\eta_{\alpha\beta}\frac{\partial \xi^{\alpha}}{\partial x^{\mu}}\frac{\partial \xi^{\beta}}{\partial x^{\nu}}dx^{\mu}dx^{\nu}=g_{\mu\nu}dx^{\mu}dx^{\nu}.
\end{equation}
 
\begin{figure}  
\hbox{
   \psfig{figure=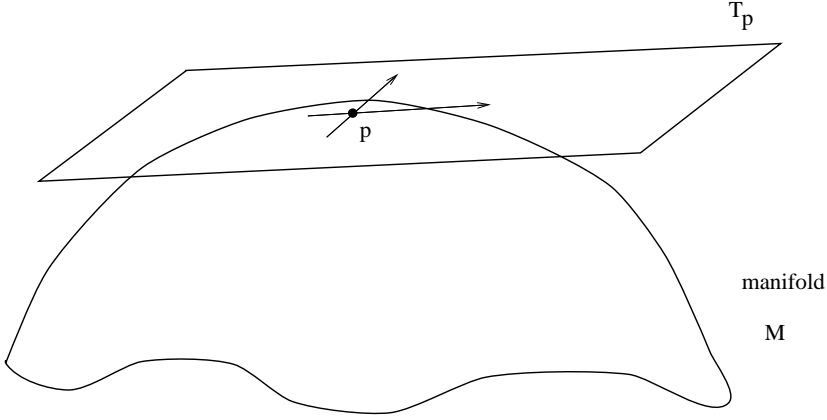,angle=-90,width=11.cm}          
        }
\caption{Tangent flat space at a point P of a curved manifold. From Carroll (2003).}\label{flat}
\end{figure}

 In the absence of gravity we can always find a global coordinate system ($\xi^{\alpha}$) for which the metric can take the form given by Eq. (\ref{eta}) everywhere. With gravity, on the contrary, such a coordinate system can represent spacetime only in an infinitesimal neighborhood of a given point. This situation is represented in Fig \ref{flat}, where the tangent flat space to a point P of the manifold is shown.   

The equation of motion of a free particle (i.e. only subject to gravity) in a general spacetime of metric $g_{\mu\nu}$ is:
\begin{equation}
	\frac{d^{2}x^{\mu}}{ds^{2}}+\Gamma^{\mu}_{\nu\rho}\frac{dx^{\nu}}{ds}\frac{dx^{\rho}}{ds}=0,\label{motion}
\end{equation}
 where $\Gamma^{\mu}_{\nu\rho}$ is the affine connection of the manifold, which is expressed in terms of derivatives of the metric tensor:
 
\begin{equation}
	\Gamma^{\mu}_{\nu\rho}=\frac{1}{2}g^{\mu\alpha}(\partial_{\nu}g_{\rho\alpha}+\partial_{\rho}g_{\nu\alpha}-\partial_{\alpha}g_{\nu\rho}).
	\end{equation}
  Here we use the convention: $\partial_{\nu} f=\partial f / \partial x^{\nu}$ and $g^{\mu\alpha}g_{\alpha\nu}=\delta^{\mu}_{\nu}$. Notice that under a coordinate transformation from $x^{\mu}$ to $x^{'\mu}$ the affine connection is not transformed as a tensor, despite that the metric $g_{\mu\nu}$ is a tensor of second rank. 
  
In a Riemannian spacetime the usual partial derivative is not a meaningful quantity since we can give it different values through different choices of coordinates. We can define a covariant differentiation through the condition of parallel transport:
\begin{equation}
	A_{\mu};_{\nu}=\frac{\partial A_{\mu}}{\partial x^{\nu}}- \Gamma^{\lambda}_{\mu\nu}A_{\lambda}.
\end{equation}
In this way, a tangent vector satisfies $V^{\nu}V_{\nu};_{\mu}=0$. If there is a vector $\zeta^{\mu}$ pointing in the direction of a symmetry of spacetime, then it can be shown (e.g. Weinberg 1972):
\begin{equation}
	\zeta_{\mu};_{\nu}+\zeta_{\nu};_{\mu}=0.
\end{equation}
This equation is called Killing's equation. A vector field $\zeta^{\mu}$ satisfying such a relation is called a Killing vector.

From Eq. (\ref{motion}) we can recover the classical Newtonian equations if: 
$$ \Gamma^{0}_{i,j}=0, \;\; \Gamma^{i}_{0,j}=0, \;\;  \Gamma^{i}_{0,0}=\frac{\partial \phi}{\partial x^{i}},$$
where $i,\;j=1,2,3$ and $\phi$ is the Newtonian gravitational potential. Then:
$$x^{0}=ct=c\tau,$$
$$\frac{d^{2}x^{i}}{d\tau^{2}}=-\frac{\partial \phi}{\partial x^{i}}.$$
We see, then, that the metric represents the gravitational potential and the affine connection the gravitational field. 

The presence of gravity is indicated by the curvature of spacetime. The Riemann tensor, or curvature tensor, provides a measure of this curvature:
\begin{equation}
	R^{\sigma}_{\mu\nu\lambda}=\Gamma^{\sigma}_{\mu\lambda, \nu}-\Gamma^{\sigma}_{\mu\nu, \lambda}+\Gamma^{\sigma}_{\alpha\nu}\Gamma^{\alpha}_{\mu\lambda}-\Gamma^{\sigma}_{\alpha\lambda}\Gamma^{\alpha}_{\mu\nu}.
\end{equation}
The Ricci tensor is defined by:
\begin{equation}
	R_{\mu\nu}=g^{\lambda\sigma}R_{\lambda\mu\sigma\nu}=R^{\sigma}_{\mu\sigma\nu}.
\end{equation}
Finally, the Ricci scalar is $R=g^{\mu\nu}R_{\mu\nu}$.

The key issue to determine the geometric structure of spacetime, and hence to specify the effects of gravity, is to find the law that fixes the metric once the source of the gravitational field is given. The source of the gravitational field is the energy-momentum tensor $T_{\mu\nu}$ that represents the physical properties of a material thing. For the simple case of a perfect fluid the energy-momentum tensor takes the form:
\begin{equation}
	T_{\mu\nu}=(\epsilon +p) u_{\mu} u_{\nu} -pg_{\mu\nu},
\end{equation}
where $\epsilon$ is the mass-energy density, $p$ is the pressure, and $u^{\mu}=dx^{\mu}/ds$ is the 4-velocity.
The field equations were found by Einstein (1915) and independently by Hilbert (1915) on November 25th and 20th, 1915, respectively. These equations can be written in the simple form:
\begin{equation}
	R_{\mu\nu}-\frac{1}{2}g_{\mu\nu}R=(8\pi G/c^{4}) T_{\mu\nu}. \label{einstein}
\end{equation}
This is a set of ten non-linear partial differential equations for the metric coefficients. The set of equations is not unique: we can add any constant multiple of $g_{\mu\nu}$ to the left member of (\ref{einstein}) and still obtain a {\sl consistent} set of equations. It is usual to denote this multiple by $\Lambda$, so the field equations can also be written as:
\begin{equation}
	R_{\mu\nu}-\frac{1}{2}g_{\mu\nu}R+\Lambda g_{\mu\nu}=(8\pi G/c^{4}) T_{\mu\nu}. \label{lambda}
\end{equation}
Lambda is a new universal constant called, because of historical reasons, the {\sl cosmological constant}. If we consider some kind of ``substance'' with equation of state given by $p=-\rho c^{2}$, then its energy-momentum tensor would be:
\begin{equation}
	T_{\mu\nu}=-p g_{\mu\nu}= \rho c^{2} g_{\mu\nu}.
\end{equation}
Notice that the energy-momentum tensor of this substance depends only on the spacetime metric $g_{\mu\nu}$, hence it describes a property of the ``vacuum'' itself. We can call $\rho$ the energy density of the vacuum field. Then, we  rewrite Eq. (\ref{lambda}) as:
\begin{equation}
	R_{\mu\nu}-\frac{1}{2}g_{\mu\nu}R=(8\pi G/c^{4}) (T_{\mu\nu}+T^{\rm vac}_{\mu\nu}), 
\end{equation}
in such a way that
\begin{equation}
	\rho_{\rm vac}c^{2}=\frac{\Lambda c^{4}}{8\pi G}.
\end{equation}
 There is evidence (e.g. Perlmutter et al. 1999) that the energy density of the vacuum is different from zero. This means that $\Lambda$ is small, but not zero. The negative pressure seems to be driving a ``cosmic acceleration''.  

The conservation of mass-energy and momentum can be derived from the field equations:
\begin{equation}
	T^{\mu\nu};_{\nu}=0. \label{conservation}
\end{equation}
 
Despite the complexity of Einstein's field equations a large number of exact solutions have been found. They are usually obtained imposing symmetries on the spacetime in such a way that the metric coefficients can be found. The first and most general solution to Eqs. (\ref{einstein}) was obtained by Karl Schwarszchild in 1916, short before he died in the Eastern Front of World War I. This solution, as we will see, describes a non-rotating black hole of mass $M$.

\section{Schwarzschild black holes}

The Schwarzschild solution for a static mass $M$ can be written in spherical coordinates $(t,r,\theta,\phi)$ as:
\begin{equation}
	ds^{2}= \left(1-\frac{2GM}{rc^{2}}\right) c^{2}dt^2- \left(1-\frac{2GM}{rc^{2}}\right)^{-1} dr^{2} -r^{2} (d\theta^{2}+\sin^{2} d\phi^{2}).\label{Schw}
\end{equation}
 This solution corresponds to the vacuum region exterior to the spherical object of mass $M$. Inside the object, spacetime will depend on the peculiarities of the physical object.   
 
 The metric given by Eq. (\ref{Schw}) has some interesting properties. Let's assume that the mass $M$ is concentrated at $r=0$. There seems to be two singularities at which the metric diverges: one at $r=0$ and the other at 
\begin{equation}
r_{\rm Schw}=\frac{2GM}{c^{2}}.	
\end{equation}
$r_{\rm Schw}$ is know as the {\sl Schwarzschild radius}. If the object of mass $M$ is macroscopic, then $r_{\rm Schw}$ is inside it, and the solution does not apply. For instance, for the Sun $r_{\rm Schw}\sim 3$ km.  However, for a point mass, the Schwarzschild radius is in the vacuum region and spacetime has the structure given by (\ref{Schw}). In general, we can write:
$$ r_{\rm Schw}\sim 3 \left( \frac{M}{M_{\odot}} \right)\;\;\; \rm km. $$

It is easy to see that strange things occur close to $ r_{\rm Schw}$. For instance, the, for the proper time we get:
\begin{equation}
	d\tau=\left(1-\frac{2GM}{rc^{2}}\right)^{1/2}\;dt. \label{time}
\end{equation}
When $r\longrightarrow \infty$ both times agree, so $t$ is interpreted as the proper time measure from an infinite distance. As the system with proper time $\tau$ approaches to $ r_{\rm Schw}$, $dt$ tends to infinity according to Eq. (\ref{time}). The object never reaches the Schwarszchild surface when seen by an infinitely distant observer. The closer the object is to the Schwarzschild radius, the slower it moves for the external observer. 

A direct consequence of the difference introduced by gravity in the local time respect to the time of an observer at infinity is that the radiation that escapes from a given $r>r_{\rm Schw}$ will be redshifted when received by a distant and static observer. Since the frequency (and hence the energy) of the photon depend of the time interval, we can write, from Eq. (\ref{time}):
\begin{equation}
	\lambda_{\infty}=\left(1-\frac{2GM}{rc^{2}}\right)^{-1/2} \lambda.
\end{equation}
Since the redshift is:
\begin{equation}
	z=\frac{\lambda_{\infty}-\lambda}{\lambda},
\end{equation}
then
\begin{equation}
	1+z=\left(1-\frac{2GM}{rc^{2}}\right)^{-1/2},
\end{equation}
and we see that when $r\longrightarrow r_{\rm Schw}$ the redshift becomes infinite. This means that a photon needs infinite energy to escape from inside the region determined by $r_{\rm Schw}$. Events that occur at $r<r_{\rm Schw}$ are disconnected from the rest of the universe. Hence, we call the surface determined by $r=r_{\rm Schw}$ an {\sl event horizon}. Whatever crosses the event horizon will never return. This is the origin of the expression ``black hole'', introduced by John A. Wheeler in the mid 1960s. The black hole is the region of spacetime inside the event horizon.  
  
What happens to an object when it crosses the event horizon?. According to Eq. (\ref{Schw}), there is a singularity at $r=r_{\rm Schw}$. However, the metric coefficients can be made regular by a change of coordinates. For instance we can consider Eddington-Finkelstein coordinates. Let us define a new radial coordinate $r_{*}$ such that radial null rays satisfy $d(t\pm r_{*})=0$. Using Eq. (\ref{Schw}) it can be shown that:
$$r_{*}=r +  \frac{2GM}{c^{2}} \log \left|\frac{r-2GM/c^{2}}{2GM/c^{2}}\right|.$$
Then, we introduce: 
$$v=ct+r_{*}.$$ The new coordinate $v$ can be used as a time coordinate replacing $t$ in Eq. (\ref{Schw}). This yields: $$ ds^{2}=\left(1-\frac{2GM}{rc^{2}}\right) (c^2dt^{2}-dr^{2}_{*})-r^{2} d\Omega^{2}$$ or
\begin{equation}
	ds^{2}=\left(1-\frac{2GM}{rc^{2}}\right) dv^{2}-2drdv -r^{2} d\Omega^{2}, \label{EF}
\end{equation}
where $$d\Omega^{2}=d\theta^{2}+\sin^{2} \theta d\phi^{2}.$$

\begin{figure}  
\hbox{
   \psfig{figure=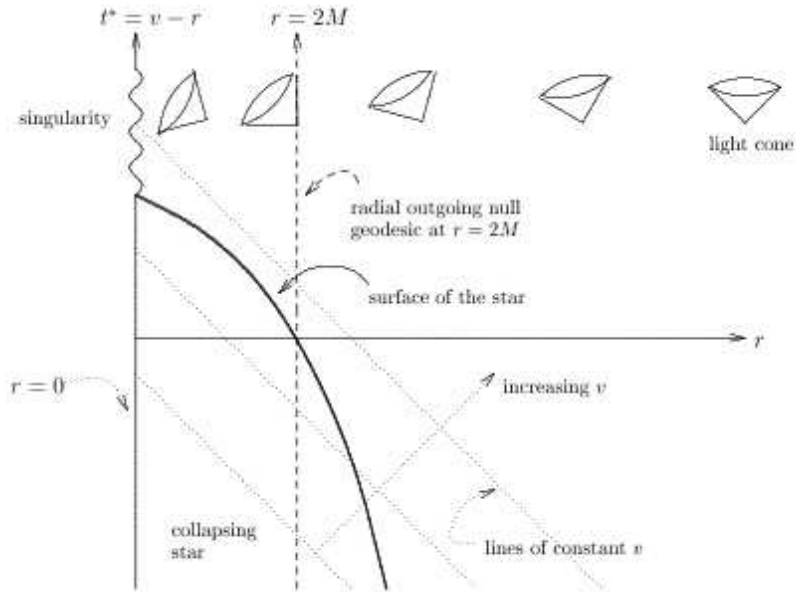}          
        }
\caption{Spacetime diagram in Eddington-Finkelstein coordinates showing an object falling down into a black hole. $r=2M=r_{\rm Schw}$ is the Schwarzschild radius where the event horizon is located (units $G=c=1$). Adapted form Townsend (1997).} \label{falling}
\end{figure}

Notice that in Eq. (\ref{EF}) the metric is non-singular at $r=2GM/c^{2}$. The only real singularity is at $r=0$, since there the Riemann tensor diverges. In order to plot the spacetime in a $(t,\; r)$-plane, we can introduce a new time coordinate $t_{*}=v-r$. From the metric (\ref{EF}) or from Fig. \ref{falling} we see that the line $r=r_{\rm Schw}$, $\theta=$constant, and $\phi=$ constant is a null ray, and hence, the surface at $r=r_{\rm Schw}$ is a null surface. This null surface is an event horizon because inside $r=r_{\rm Schw}$ all cones have $r=0$ in there future. The object in $r=0$ is the source of the gravitational field and is called {\sl the singularity}. We will say more about it in Sect. \ref{sing} For the moment, we only remark that everything that crosses the event horizon will end at the singularity. This is the inescapable fate. There is no way to avoid it: in the future of every event inside the event horizon is the singularity. There is no escape, no hope, no freedom, inside the black hole. There is just the singularity, whatever such a thing might be. 

We see now that the name ``black hole'' is not strictly correct for spacetime regions isolated by event horizons. There is no hole to other place. What falls into the black hole, goes to the singularity. The singularity increases its mass and energy, and then the event horizon grows. This would not happen if what falls into the hole were able to pass through, like through a hole in the wall. A black hole is more like a spacetime precipice, deep, deadly, and with something unknown at the bottom.  

Orbits around a Schwazschild black hole can be easily calculated using the metric and the relevant symmetries (see. e.g. Raine and Thomas 2005). For circular orbits of a massive particle we have the conditions
$$\frac{dr}{d\tau}=0, \;\;\; {\rm and} \;\;\; \frac{d^{2}r}{d\tau^{2}}=0.$$ The orbits are possible only at the turning points of the effective potential:
\begin{equation}
	V_{\rm eff}=\sqrt{\left(1+\frac{L^{2}}{r^{2}}\right)\left(1-\frac{2r_{\rm g}}{r}\right)},
\end{equation}
where $L$ is the angular momentum in units of $m_0 c$ and $r_{\rm g}=GM/c^{2}$ the gravitational radius. Then,
\begin{equation}
	r=\frac{L^{2}}{2r_{\rm g}}\pm\frac{1}{2}\sqrt{\frac{L^{4}}{r^{2}_{\rm g}}-12L^{2}}.
\end{equation}
For $L^{2}>12 r_{\rm g}^{2}$ there are two solutions. The negative sign corresponds to a maximum of the potential and is unstable, and the positive sign corresponds to minimum, which is stable. At $L^{2}=12 r_{\rm g}^{2}$ there is a single stable orbit. It is the innermost marginally stable orbit, and it occurs at $r= 6 r_{\rm g}=3 r_{\rm Schw}$. The specific angular momentum of a particle in a circular orbit is:
$$L=\left(\frac{r_{\rm g} r}{1-3r_{\rm g}/r}\right)^{1/2}.$$ Its energy (units of $m_0 c^{2}$) is: 
$$E=\left(1-\frac{2r_{\rm g}}{r}\right)\left(1-\frac{3r_{\rm g}}{r}\right)^{-1/2}.$$ The proper and observer's periods are: 
$$\tau=\frac{2\pi}{c}\left(\frac{r^{3}}{r_{\rm g}}\right)^{1/2}\left(1-\frac{3r_{\rm g}}{r}\right)^{1/2}$$ and 
$$ T= \frac{2\pi}{c} \left(\frac{r^{3}}{r_{\rm g}}\right)^{1/2}.$$ Notice that when $r\longrightarrow 3r_{\rm g}$ both $L$ and $E$ tend to infinity, so only massless particles can orbit at such a radius. 

The local velocity at $r$ of an object falling from rest to the black hole is (e.g. Raine and Thomas 2005):
\begin{equation}
	v_{\rm loc}=\left(\frac{2r_{\rm g}}{r}\right)^{1/2} \;\;\; ({\rm in\; units\; of}\; c).
\end{equation}
Then, the differential acceleration the object will experience along an element $dr$ is\footnote{Notice that $dv_{\rm loc}/d\tau=(dv_{\rm loc}/dr)(dr/d\tau)=(dv_{\rm loc}/dr)v_{\rm loc}=r_{\rm g} c^{2}/r^{2}$.}:
\begin{equation}
	dg=\frac{2r_{\rm g}}{r^{3}} c^{2}\;dr.
\end{equation}
The tidal acceleration on a body of finite size $\Delta r$ is simply $2r_{\rm g}/r^{3} c^{2}\;\Delta r$.  This acceleration and the corresponding force becomes infinite at the singularity. As the object falls into the black hole, tidal forces act to tear it apart. This painful process is known as ``spaghettification''. The process can be significant long before crossing the event horizon, depending on the mass of the black hole.  

The energy of a particle in the innermost stable orbit can be obtained from the above equation for the energy setting $r=6 r_{\rm g}$. This yields:
$$E=\left(1-\frac{2r_{\rm g}}{6r_{\rm g}}\right)\left(1-\frac{3r_{\rm g}}{6r_{\rm g}}\right)^{-1/2}=\frac{2}{3}\sqrt{2}.$$
Since a particle at rest at infinity has $E=1$, then the energy that the particle should release to fall into the black hole is $1-(2/3)\sqrt{2}=0.057$. This means 5.7 \% of its rest mass energy, significantly higher than the energy release that can be achieved through nuclear fusion.

An interesting question we can make is what is the gravitational acceleration at the event horizon as seen by an observer from infinity. The acceleration relative to a hovering frame system of a freely falling object at rest at $r$ is (Raine and Thomas 2005): 
$$g_{r}=- c^{2}\left(\frac{GM/c^{2}}{r^{2}}\right)\left(1-\frac{2GM/c^{2}}{r}\right)^{-1/2}.$$ So, the energy spent to move the object a distance $dl$ will be $dE_{r}=m g_{r} dl$. The energy expended respect to a frame at infinity is 
$dE_{\infty}=m g_{\infty} dl$. Because of the conservation of energy, both quantities should be related by a redshift factor:
$$\frac{E_{r}}{E_{\infty}}=\frac{g_{r}}{g_{\infty}}=\left(1-\frac{2GM/c^{2}}{r}\right)^{-1/2}.$$ Hence, using the expression for $g_{r}$ we get:
\begin{equation}
	g_{\infty}=c^{2}\frac{GM/c^{2}}{r^{2}}.
\end{equation} 
Notice that for an observer at $r$, $g_{r}\longrightarrow\infty$ when $r\longrightarrow r_{\rm Schw}$. However, from infinity the required force to hold the object hovering at the horizon is:
$$m g_{\infty}= c^{2}\frac{GmM/c^{2}}{r_{\rm Schw}^{2}}=\frac{m c^{4}}{4GM}.$$ This is the {\sl surface gravity} of the black hole. 

Other coordinates can be introduced to study additional properties of black holes. We refer the reader to the books of Frolov and Novikov (1998) and Raine and Thomas (2005) for Kruskal coordinates and further details. Now, we turn to axially symmetric (rotating) solutions of the field equations.

\section{Kerr black holes}

A Schwarzschild black hole does not rotate. The solution of the field equations (\ref{einstein}) for a rotating body of mass $M$ and angular momentum per unit mass $a$ was found by Roy Kerr (1963):

\begin{eqnarray} ds^2&=&g_{tt}dt^2+2g_{t\phi}dtd\phi-g_{\phi\phi}d\phi^2-\Sigma\Delta^{-1}dr^2-\Sigma d\theta^2\label{Kerr}\\ g_{tt}&=& (c^2-2GMr\Sigma^{-1})
\label{Kerr1}\\ g_{t\phi}&=&2GMac^{-2}\Sigma^{-1}r\sin^2\theta\\ g_{\phi\phi}&=&[(r^2+a^2c^{-2})^2 -a^2c^{-2}\Delta \sin^2\theta ]\Sigma^{-1}\sin^2\theta\\\Sigma&\equiv& r^2+a^2 c^{-2}\cos^2\theta\\\Delta&\equiv& r^2-2GMc^{-2}r+a^2c^{-2}.
\label{Kerr2}\end{eqnarray} 
This is the Kerr metric in Boyer-Lindquist coordinates $(t,\;r,\;\theta,\;\phi)$, which reduces to Schwarzschild metric for $a=0$. In Boyer-Lindquist coordinates the metric is approximately Lorentzian at infinity (i.e. we have a Minkowski spacetime in in the usual coordinates of Special Relativity).

Note that the element $g_{t\phi}$ no longer vanishes. Even at infinity this element remains (hence we wrote {\sl approximately} Lorentzian above). The Kerr parameter $a c^{-1}$ has dimensions of length. The larger the ratio of this scale to $GMc^{-2}$ (the {\it spin\ parameter\/} $a_*\equiv ac/GM$), the more aspherical the metric. Schwarzschild's black hole is the special case of Kerr's for $a=0$. Notice that, with the adopted conventions, the angular momentum $J$ is related to the parameter $a$ by:
\begin{equation}
	J=Ma.
\end{equation}

Just as the Schwarzschild solution is the unique static vacuum solution of Eqs. (\ref{einstein}) (a result called Israel's theorem), the Kerr metric is the unique stationary axisymmetric vacuum solution (Carter-Robinson theorem).  
  
The horizon, the surface which cannot be crossed outward, is determined by the condition $g_{rr}\rightarrow\infty$ ($\Delta=0$).  It lies at $r=r_h$ where
\begin{equation}  
r_h \equiv GMc^{-2}+[(GMc^{-2})^2-a^2c^{-2}]^{1/2}.\label{rh}
\end{equation} 
Indeed, the track $r=r_h$, $\theta=$ constant with $d\phi/d\tau=a(r_h{}^2+a^2)^{-1}\, dt/d\tau$ has $d\tau=0$ (it represents a photon circling azimuthaly {\it on\/} the horizon, as opposed to hovering at it).  Hence the surface $r=r_h$ is tangent to the local light cone.  Because of the square root in Eq. (\ref{rh}), the horizon is well defined only for $a_*= ac/GM \leq 1$. An {\sl extreme} (i.e. maximally rotating) Kerr black hole has a spin parameter $a_{*}=1$. Notice that for $(GMc^{-2})^2-a^2c^{-2}>0$ we have actually two horizons. The second, the {\sl inner} horizon, is located at:
\begin{equation}  
r^{\rm inn}_h \equiv GMc^{-2}-[(GMc^{-2})^2-a^2c^{-2}]^{1/2}.\label{rh_inn}
\end{equation}
This horizon is not seen by an external observer, but it hides the singularity to any observer that has already crossed $r_{h}$ and is separated from the rest of the universe. For $a=0$, $r^{\rm inn}_{h}=0$ and $r_h =r_{\rm Schw}$. The case $(GMc^{-2})^2-a^2c^{-2}<0$ corresponds to no horizons and it is thought to be unphysical. 

\begin{figure}  
\includegraphics[width=5cm]{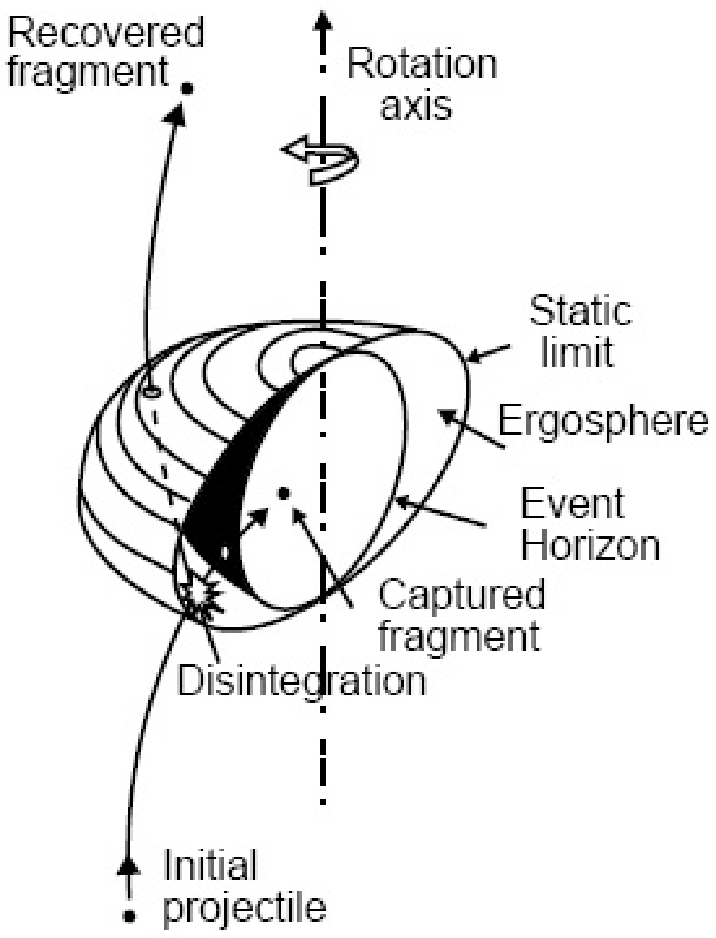}
\includegraphics[trim=0 -50 0 0,clip,width=8cm]{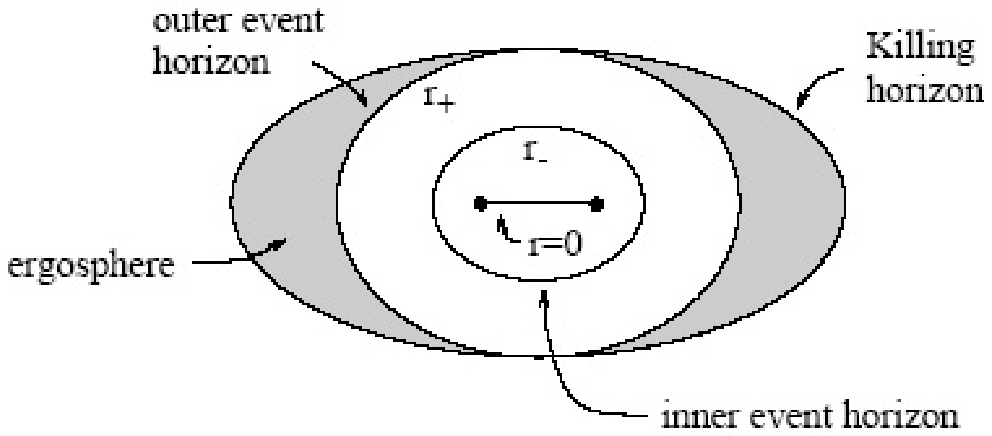}

\caption{{Left}: A rotating black hole and the Penrose process. Adapted from J-P. Luminet (1998). {\sl Right}: Sketch of the interior of a Kerr black hole.} \label{ergo}
\end{figure}

A study of the orbits around a Kerr black hole is beyond the limits of the present article (the reader is referred to Frolov and Novikov 1998), but we will mention several interesting features. One is that if a particle initially falls radially with no angular momentum from infinity to the black hole, it gains angular motion during the infall. The angular velocity as seen from a distant observer is:
\begin{equation}
	\Omega (r,\; \theta)= \frac{d\phi}{dt}=\frac{2GM/c^{2}a r}{(r^{2}+a^{2}c^{-2})^{2}-a^{2}c^{-2}\Delta\sin ^{2}\theta}.
\end{equation}
Notice that the particle will acquire angular velocity in the direction of the spin of the black hole. As the black hole is approached, the particle will find an increasing tendency to get carried away in the same sense in which the black hole is rotating. To keep the particle stationary respect the distant stars, it will be necessary to apply a force against this tendency. The closer the particle will be to the black hole, the stronger the force. At a point $r_{\rm e}$ it becomes impossible to counteract the rotational sweeping force. The particle is in a kind of spacetime maelstrom. The the surface determined by $r_{\rm e}$ is the {\sl static limit}: from there in, you cannot avoid to rotate. Spacetime is rotating here in such a way that you cannot do anything in order to not co-rotate. You can still escape from the black hole, since the outer event horizon has not been crossed, but rotation is inescapable. The region between the static limit and the event horizon is called the {\sl ergosphere}. The ergosphere is not spherical but its shape changes with the latitude $\theta$:
\begin{equation}
	r_{\rm e}=\frac{GM}{c^{2}}+\frac{1}{c^{2}}\left(G^{2}M^{2}-a^{2}\cos^{2}\theta\right)^{1/2}.
\end{equation}
 The static limit lies outside the horizon except at the poles where both surfaces coincide. The phenomenon of ``frame dragging''' is common to all axially symmetric metrics with $d_{t\phi}\neq 0$   

Roger Penrose (1969) suggested that a projectile thrown from outside into the ergosphere begins to rotate acquiring more rotational energy than it originally had. Then the projectile can break up into two pieces, one of which will fall into the black hole, whereas the other can go out of the ergosphere. The piece coming out will then have more energy than the original projectile. In this way, we can extract energy from a rotating black hole. In Fig. \ref{ergo} we illustrate the situation and show the static limit, the ergosphere and the outer/inner horizons of a Kerr black hole.  

The innermost marginally stable circular orbit $r_{\rm ms}$ around a extreme rotating black hole ($ac^{-1}=GM/c^{2}$) is given by (Raine and Thomas 2005):
\begin{equation}
	\left(\frac{r_{\rm ms}}{GM/c^{2}}\right)^{2}-6\left(\frac{r_{\rm ms}}{GM/c^{2}}\right)\pm 8 \left(\frac{r_{\rm ms}}{GM/c^{2}}\right)^{1/2}-3=0.
\end{equation}
For the $+$ sign this is satisfied by $r_{\rm ms}= GM/c^{2}$, whereas for the $-$ sign the solution is $r_{\rm ms}=9 GM/c^{2}$. The first case corresponds to a co-rotating particle and the second one to a counter-rotating particle. The energy of the co-rotating particle in the innermost orbit is $1/\sqrt{3}$ (units of $mc^{2}$). The binding energy of a particle in an orbit is the difference between the orbital energy and its energy at infinity. This means a binding energy of 42\% of the rest energy at infinity!. For the counter-rotating particle, the binding energy is 3.8 \%, smaller than for a Schwazschild black hole.  

\section{Other black holes}

\subsection{Reissner-Nordstr{\o}m black holes}

The Reissner-Nordstr{\o}m metric is a spherically symmetric solution of Eqs. (\ref{einstein}). However, it is not a vacuum solution, since the source has an electric charge $Q$, and hence there is an electromagnetic field. The energy-momentum tensor of this field is:
\begin{equation}
	T_{\mu\nu}=-\mu_{0}^{-1}(F_{\mu\rho}F_{\nu^{\rho}}-\frac{1}{4} g_{\mu\nu} F_{\rho\sigma}F^{\rho\sigma}),
\end{equation}
where $F_{\mu\nu}=\partial_{\mu} A_{\nu}-\partial_{\nu}A_{\mu}$ is the electromagnetic field strength tensor and $A_{\mu}$ is the electromagnetic 4-potential. Outside the charged object the 4-current $j^{\mu}$ is zero, so the Maxwell equations are:
\begin{eqnarray}
	F^{\mu\nu};\mu&=&0,\\
	F_{\mu\nu};\sigma+F_{\sigma\mu};\nu+F_{\nu\sigma};\mu&=&0.
\end{eqnarray}
The Einstein and Maxwell equations are coupled since $F^{\mu\nu}$ enters into the gravitational field equations through the energy-momentum tensor and the metric $g_{\mu\nu}$ enters into the electromagnetic equations through the covariant derivative. Because of the symmetry constraints we can write:
\begin{equation}
	[A^{\mu}]=\left(\frac{\varphi(r)}{c^{2}},\;a(r),\;0\;0\right),
\end{equation}
 where $\varphi(r)$ is the electrostatic potential, and $a(r)$ is the radial component of the 3-vector potential as $r\longrightarrow \infty$. 

The solution for the metric is given by 
\begin{equation}
  ds^2 = \Delta c^{2}dt^2 - \Delta^{-1} dr^2 -
  r^2d\Omega^2\ ,\label{RNBH}
\end{equation}
where
\begin{equation}
  \Delta = 1-\frac{2GM/c^{2}}{r}+\frac{q^2}{r^2}.
  \label{Q}
\end{equation}
In this expression, $M$ is once again interpreted as 
the mass of the hole and 
\begin{equation}
q=\frac{GQ^{2}}{4\pi \epsilon_{0} c^{4}} \label{q}
\end{equation} 
is related to the total electric charge $Q$.

The metric has a coordinate singularity at $\Delta=0$, in such a way that:
\begin{equation}
	r_{\pm}=r_{\rm g}\pm (r_{\rm g}^{2}-q^{2})^{1/2}.
\end{equation}
 Here, $r_{\rm g}=GM/c^{2}$ is the gravitational radius. For $r_{\rm g}=q$, we have an {\sl extreme} Reissner-Nordstr{\o}m black hole with a unique horizon at $r=r_{\rm g}$. For the case $r_{\rm g}^{2}>q^{2}$, both $r_{\pm}$ are real and there are two horizons. Finally, in the case $r_{\rm g}^{2}<q^{2}$ both $r_{\pm}$ are imaginary there is no coordinate singularities, no horizon hides the intrinsic singularity at $r=0$. It is thought that naked singularities do not exist in nature (see Section \ref{sing} below).  
 
\subsection{Kerr-Newman black holes}

The Kerr-Newman metric of a charged spinning black hole is the most general black hole solution. This metric can be obtained from the Kerr metric (\ref{Kerr}) in Boyer-Lindquist coordinates by the replacement:
$$\frac{2GM}{c^{2}} r \longrightarrow  \frac{2GM}{c^{2}} r -q^{2}, $$ where $q$ is related to the charge $Q$ by Eq. (\ref{q}). The Kerr-Newman solution is a non-vacuum solution, as the Reissner-Nordstr{\o}m is. It shares with Kerr and Reissner-Nordstr{\o}m solutions the existence of two horizons, and as the Kerr solution it presents an ergosphere. Since the electric field cannot remain static in the ergosphere, a magnetic field is generated as seen by an observer outside the static limit. Charged black holes might be a natural result from charge separation during the gravitational collapse of a star. It is thought that an astrophysical charged object would discharge quickly by accretion of charges of opposite sign. However, there remains the possibility that the charge separation could lead to a configuration where the black hole has a charge and a superconducting ring around it would have the same but opposite charge, in such a way the whole system seen from infinity is neutral. In such a case a Kerr-Newman black hole might survive for some time, depending on the environment. For details, the reader is referred to the highly technical book by Brian Punsly (2001).  

\section{Black hole formation}

Black holes will form every time that matter is compressed beyond its Schwarzschild radius. This can occur in a variety of forms, from particle collisions to the implosion of stars or the collapse of dark matter in the early universe. The most common black hole formation mechanism in our Galaxy seems to be gravitational collapse. A normal star is stable as long as the nuclear reactions occurring in its interior provide thermal pressure to support it against gravity. Nuclear burning gradually transforms the stellar core from H to He and in the case of massive stars ($M>5\;M_{\odot}$) then to C and finally to Fe. The core contracts in the processes, in order to achieve the ignition of each phase of thermonuclear burning. 

Finally, the endothermic disintegration of iron-group nuclei, which are those with the tightest bound, precipitates the collapse of the core to a stellar-mass black hole. Stars with masses in the range $20-30 \;M_{\odot}$ produce black holes with $M >1.8\;M_{\odot}$. Low-mass black holes ($1.5 \;M_{\odot}< M <1.8\;M_{\odot}$) can result from the collapse of stars of $18-20 \;M_{\odot}$ along with the ejection of the outer layers of the star by a shock wave in an event known as Type II supernova. A similar event, occurring in stars with $10-18 \;M_{\odot}$ leaves behind a neutron star. Very massive stars with high spin likely end producing a gamma-ray burst and a very massive ($M>10\;M_{\odot}$) black hole. The binary stellar systems have a different evolution. The interested reader can find an comprehensive review in Brown et al. (1999). 

\begin{figure}  
\includegraphics[trim=0 10 0 0,clip,width=13cm]{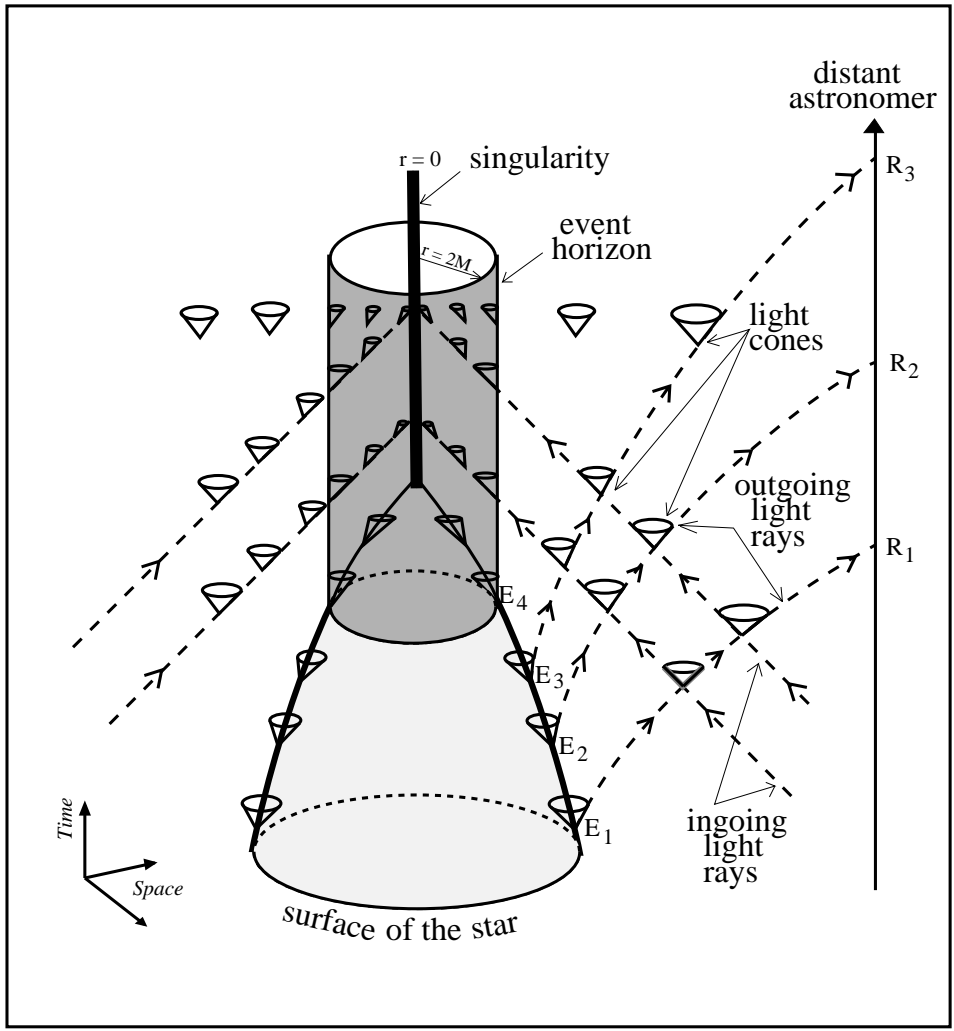}

\caption{An Eddington-Finkelstein diagram of a collapsing star with the subsequent black hole formation. Adapted from J-P. Luminet (1998).} \label{collapse}
\end{figure} 

In Fig. \ref{collapse} we show the Eddignton-Finkelstein diagram of the gravitational collapse of a star. Once the null surface of the light cones points along the time axis the black hole is formed: light rays will never be again able to escape to the outer universe. 

If the collapse is not perfectly symmetric, any asymmetry in the resulting black hole is radiated away as gravitational waves, in such a way that the final result is a black hole that is completely characterized by the three parameters $M$, $J$, and $Q$. The black hole, then, has no information about the details of the formation process.       

\section{Black hole thermodynamics}

The area of a Schwarzschild black hole is:
\begin{equation}
	A_{\rm Schw}=4\pi r^{2}_{\rm Schw}=\frac{16 \pi G^{2} M^{2}}{c^{4}}. \label{A_s}
\end{equation}
In the case of a Kerr-Newman black hole, the area is:
\begin{eqnarray}
	A_{\rm KN}&=&4\pi\left(r^{2}_{+}+\frac{a^{2}}{c^{2}}\right) \nonumber\\ &=&4\pi\left[\left(\frac{GM}{c^{2}}+\frac{1}{c^{2}}\sqrt{G^{2}M^{2}-GQ^{2}-a^{2}}\right)^{2}+\frac{a^{2}}{c^{2}}\right].
\label{A_kn}
\end{eqnarray}
Notice that expression (\ref{A_kn}) reduces to (\ref{A_s}) for $a=Q=0$.

When a black hole absorbs a mass $\delta M$, its mass increases to $M+ \delta M$, and hence, the area increases as well. Since the horizon can be crossed in just one direction the area of a black hole can only increase. This suggests an analogy with entropy. A variation in the entropy of the black hole will be related to the heat ($\delta Q$) absorbed through the following equation:
\begin{equation}
	\delta S=\frac{\delta Q}{T_{\rm BH}}=\frac{\delta M c^{2}}{T_{\rm BH}}.
\end{equation}

Particles trapped in the black hole will have a wavelength:
\begin{equation}
	\lambda=\frac{\hbar c}{kT}\propto r_{\rm Schw},
\end{equation}
where $k$ is the Boltzmann constant.
Then, 
$$\xi \frac{\hbar c}{kT}=\frac{2GM}{c^{2}}, $$
where $\xi$ is a numerical constant. Then,
$$T_{\rm BH}=\xi\frac{\hbar c^{3}}{2GkM},\;\;\;{\rm and}\;\;\;S=\frac{c^{6}}{32 \pi G^{2} M}
\int\frac{dA_{\rm Schw}}{T_{\rm BH}}=\frac{c^{3} k}{16 \pi \hbar G \xi} A_{\rm Schw}+ \;{\rm constant}.$$
 
A quantum mechanical calculation of the horizon temperature in the Schwarzschild case leads to $\xi=(4\pi)^{-1}$ and hence:
\begin{equation}
	T_{\rm BH}=\frac{\hbar c^{3}}{8GMk}\cong 10^{-7}{\rm K} \left(M_{\odot}\over M \right). \label{T}
\end{equation}
 We can write then the entropy of the black hole as:
\begin{equation}
	S=\frac{kc^{3}}{4\pi \hbar G} A_{\rm Schw} + \;{\rm constant}\sim 10^{77} \left(\frac{M}{M_{\odot}}\right)^{2} k \; {\rm JK}^{-1}. \label{ent}
\end{equation}
The formation of a black hole implies a huge increase of entropy.  Just to compare, a solar mass star has an entropy $\sim 20$ orders of magnitude lower. This tremendous increase of entropy is related to the loss of information about the original system once the black hole is formed. 

The analogy between area and entropy allows to state a set of laws for black holes thermodynamics:
\begin{itemize}
	\item First law (energy conservation): $dM=T_{\rm BH}dS+ \Omega_{+} dJ +\Phi dQ$. Here, $\Omega$ is the angular velocity and $\Phi$ the electrostatic potential. 
	\item Second law (entropy never decreases): In all physical processes involving black holes the total surface area of all the participating black holes can never decrease. 
	\item Third law (Nernst's law): The temperature (surface gravity) of a black black hole cannot be zero. Since $T_{\rm BH}=0$ with $A\neq0$ for extremal charged and extremal Kerr black holes, these are thought to be limit cases that cannot be reached in Nature. 
	\item Zeroth law (thermal equilibrium): The surface gravity (temperature) is constant over the event horizon of a stationary axially symmetric black hole.    
	
\end{itemize}

\section{Quantum effects in black holes}

If a temperature can be associated with black holes, then they should radiate as any other body. The luminosity of a Schwarzschild black hole is:
\begin{equation}
	L_{\rm BH}=4\pi r_{\rm Schw}^{2} \sigma T_{\rm BH}^{4}\sim \frac{16 \pi \sigma \hbar^{4} c^{6}}{(8\pi)^{4} G^{2} M^{2} k^{4}}.
\end{equation}
Here, $\sigma$ is the Stephan-Boltzmann constant. This expression can be written as:
\begin{equation}
	L_{\rm BH}=10^{-17}\left(\frac{M_{\odot}}{M}\right)^{2}\;\;\;{\rm erg\;s}^{-1}.
\end{equation}
 The lifetime of a black hole is:
\begin{equation}
\tau\cong\frac{M}{dM/dt}\sim 2.5 \times 10^{63} \left(\frac{M}{M_{\odot}}\right)^{3}\;\;\;{\rm years}. \label{age}	
\end{equation}
Notice that the black hole heats up as it radiates!. This occurs because when the hole radiates, its mass decreases and then according to Eq. (\ref{T}) the temperature must rise. 

If nothing can escape from black holes because of the existence of the event horizon, what is the origin of this radiation?. The answer, found by Hawking (1974), is related to quantum effects close to the horizon. According to the Heisenberg relation $\Delta t \Delta E \geq \hbar/2$ particles can be created out of the ground state of a quantum field as far as the relation is not violated. Particles must be created in pairs, along a tiny time, in order to satisfy conservation laws other than energy. If a pair is created close to the horizon an one particle crosses it, then the other particle can escape provided its momentum is in the outward direction. The virtual particle is then transformed into a real particle, at expense of the black hole energy. The black hole then will lose energy and its area will decrease slowly, violating the second law of thermodynamics. However, there is no violation if we consider a {\sl generalized second law}, that always holds: {\sl In any process, the total generalized entropy $S+S_{\rm BH}$ never decreases} (Bekenstein 1973).           

\section{Black hole magnetospheres}

In the real universe black holes are not expected to be isolated, hence the ergosphere should be populated by charged particles. This plasma would rotate in the same sense as the black hole due to the effects of the frame dragging. A magnetic field will develop and will rotate too, generating a potential drop that might accelerate particles up to relativistic speed and produce a wind along the rotation axis of the hole. Such a picture has been consistently developed by Punsly and Coroniti (1990a, b) and Punsly (2001).    

Blandford and Znajek (1977) developed a general theory of force-free steady-state axisymmetric magnetosphere of a rotating black hole. In an accreting black hole, a magnetic field can be sustained by external currents, but as such currents move along the horizon, the field lines are usually representing as originating from the horizon and then being torqued by rotation. The result is an outgoing electromagnetic flux of energy and momentum. This picture stimulated the development of the so-called ``membrane paradigm'' by Thorne et al. (1986) where the event horizon is attributed with a set of physical properties. This model of black hole has been subjected to strong criticism by Punsly (2001) since General Relativity implies that the horizon is causally disconnected from the outgoing wind. 

Recent numerical simulations (e.g. Komissarov 2004) show that the key role in the electrodynamic mechanisms of rotating black holes is played by the ergosphere and not by the horizon. However, globally the Blandford-Znajek solution seems to be asymptotically stable. The controversy still goes on and a whole bunch of new simulations are exploring the different aspects of relativistic magneto-hydrodynamic (MHD) outflows from black hole magnetospheres.       

\section{Singularities}\label{sing}

A singularity is a region where the manifold $M$ that represents spacetime is {\sl incomplete}. A manifold is incomplete if it contains at least one {\sl inextendible} curve. A curve $\gamma:[0,a)\longrightarrow M$ is inextendible if there is no point $p$ in $M$ such that $\gamma(s)\longrightarrow p$ as $a\longrightarrow s$, i.e. $\gamma$ has no endpoint in $M$. A given spacetime $(M, g_{ab})$ has an {\sl extension} if there is an isometric embedding $\theta: M\longrightarrow M^{\prime}$, where $(M^{\prime}, g_{ab}^{\prime})$ is a spacetime and $\theta$ is onto a proper subset of $M^{\prime}$. A spacetime is {\sl singular} if it contains a curve $\gamma$ that is inextendible in the sense given above. Singular spacetimes contain singularities. 

A so-called {\sl coordinate singularity} is not as real singularity. It seems to be singular in some spacetime representation but it can be removed by a coordinate change, like the ``Schwarzschild singularity'' at $r_{\rm Scwh}=2GM/c^{2}$ in a Schwarzchild spacetime. We can change to Eddington-Finkelstein coordinates, for instance, and then we see that geodesic lines can go through that point of the manifold. Essential singularities cannot be removed in this way. This occurs, for instance, with the singularity at $r=0$ in the Schwarzschild spacetime or with the ring singularity at $r=0$ and $\theta=\pi/2$ in the Kerr metric written in Boyer-Lindquist coordinates\footnote{In Cartesian coordinates the Kerr singularity occurs at $x^{2}+y^{2}=a^{2}c^{-2}$ and $z=0$.}. In such cases, the curvature scalar $R^{\mu\nu\rho\sigma}R_{\mu\nu\rho\sigma}$ diverges. There is no metric there, and the Einstein equations cannot be defined. 

An essential or true singularity should not be interpreted as a representation of a physical object of infinite density, infinite pressure, etc. Since the singularity does not belong to the manifold that represents spacetime in General Relativity, it simply cannot be described or represented in the framework of such a theory. General Relativity is incomplete in the sense that it cannot provide a full description of the gravitational behavior of any physical system. True singularities are not within the range of values of the bound variables of the theory: they do not belong to the ontology of a world that can be described with 4-dimensional differential manifolds. 

Singularities do not belong to classical spacetime. This is not surprising since singularities are extremely compact systems. At such small scales, relations among things should be described in a quantum mechanical way. If spacetime is formed by the events that occur to things, it should be represented through quantum mechanic theory when the things are described by a quantum theory. Since even in the standard quantum theory time appears as a continuum variable, a new approach is necessary.     

Spacetime singularities are expected to be covered by horizons. Although formation mechanisms for naked singularities have been proposed, the following conjecture is usually considered valid:
\begin{itemize}
	\item Cosmic Censorship Conjecture (Roger Penrose): Singularities are always hidden behind event horizons.
\end{itemize}
         
The classical references on singularities are Hawking and Ellis (1973) and Clarke (1993).

\section{Wormholes}

\begin{figure}  
\includegraphics[width=6cm]{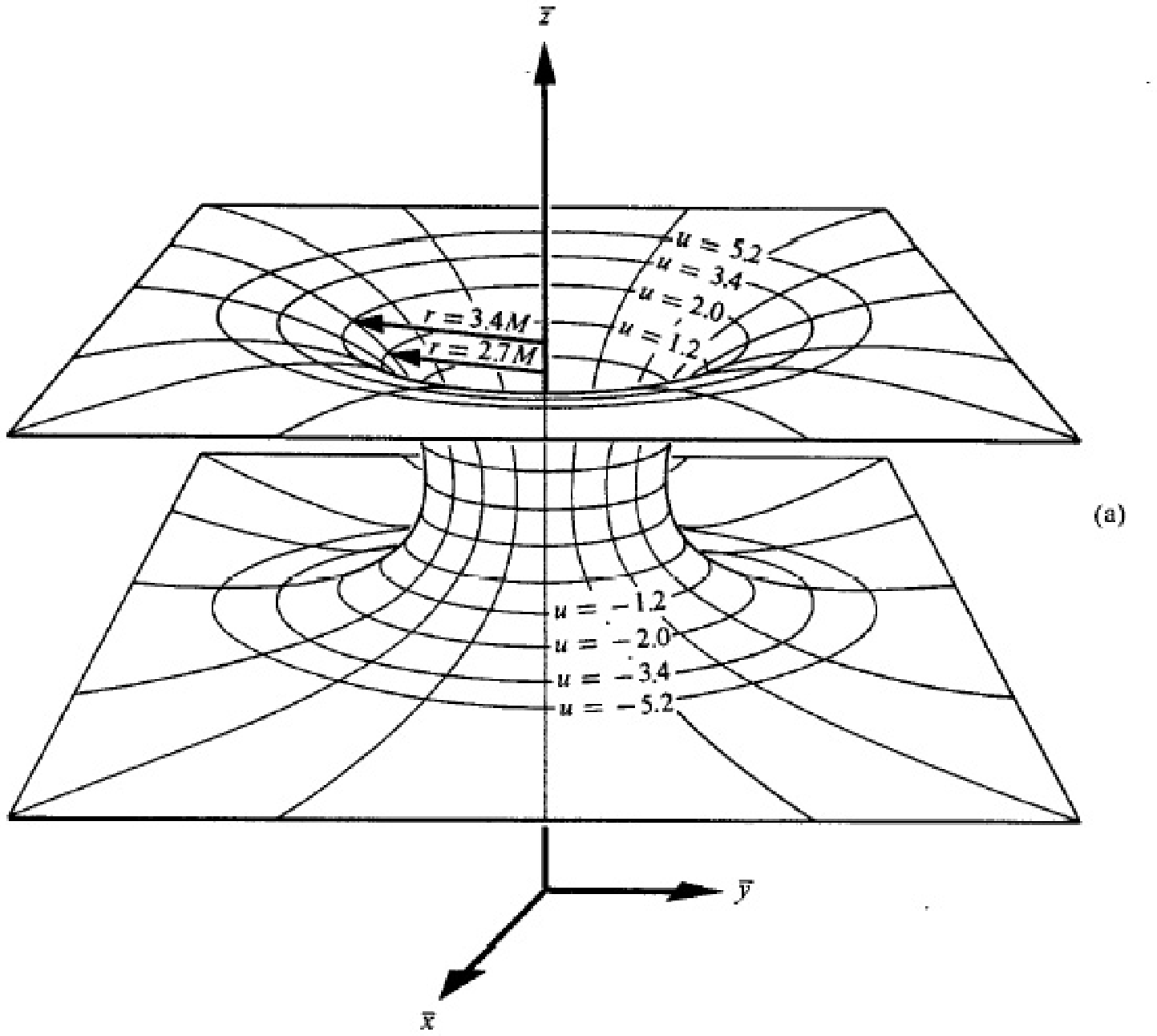}
\includegraphics[trim=0 -30 0 0,clip,width=8cm]{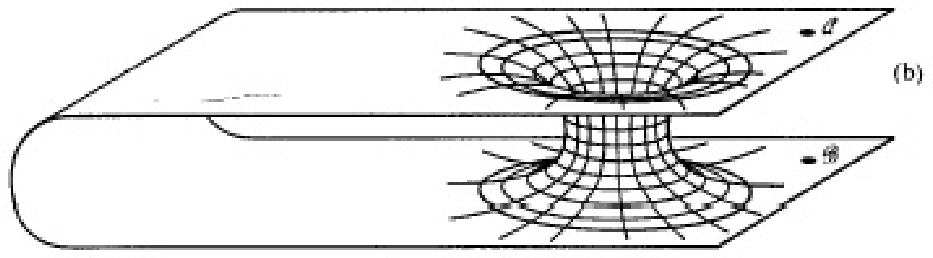}

\caption{Embedding diagrams of wormholes. Adapted from Misner et al. (1973).}\label{worm}
\end{figure}

A wormhole is a region of spacetime with non-trivial topology. It
has two mouths connected by a throat (see Figure \ref{worm}). The mouths are not hidden by
event horizons, as in the case of black holes, and, in addition,
there is no singularity to avoid the passage of particles,
or travelers, from one side to the other. Contrary to black holes, wormholes are holes in spacetime, i.e. their existence implies a multiple-connected spacetime.

There are many types of wormhole solutions for the Einstein field equations (see Visser 1996). Let us consider the static spherically symmetric line element,

$$ ds^2 =  e^{2\Phi(l)} c^{2}dt^2 - dl^2 - r(l)^2 d\Omega^2$$ 
where $l$ is a proper radial distance that covers the entire range
$(-\infty,\infty)$. In order to have a wormhole which is
traversable in principle, we need to demand that:

\begin{enumerate}

\item  $\Phi(l)$ be finite everywhere, to be consistent with the
absence of event horizons.

\item In order for the spatial geometry to tend to an appropriate asymptotically
flat limit, it must happen that $$\lim_{r\rightarrow \infty}
r(l)/l=1$$ and $$ \lim_{r\rightarrow \infty} \Phi(l)=\Phi_0<\infty
.$$
\end{enumerate}
The radius of the wormhole is defined by $r_0=\min \{ r(l) \}$, where we
can set $l=0$.

To consider wormholes which can be traversable in practice, we
should introduce additional engineering constraints. Notice that for simplicity we have considered both
asymptotic regions as interchangeable. This is the best choice of
coordinates for the study of wormhole geometries because
calculations result considerably simplified. In general, two patches are
needed to cover the whole range of $l$, but this is not noticed if
both asymptotic regions are assumed similar. The static line
element is:

\begin{equation} 
ds^2 = e^{2\Phi(r)}c^{2} dt^2 - e^{2\Lambda(r)} dr^2 - r^2
d\Omega^2,
\end{equation}
where the redshift function $\Phi$ and the
shape-like function $e^{2\Lambda}$ characterize the wormhole
topology. They must satisfy:

\begin{enumerate}

\item $e^{2\Lambda} \geq 0$
throughout the spacetime. This is required to ensure the
finiteness of the proper radial distance defined by $dl = \,\pm
\,e^\Lambda \,dr$. The $\pm$ signs refer to the two asymptotically
flat regions which are connected by the wormhole throat.

\item  The precise
definition of the wormhole's throat (minimum radius, $r_{\rm th}$) entails a
vertical slope of the embedding surface

\begin{equation} \lim_{r \rightarrow r_{\rm th}^+} \frac{dz}{dr} \, = \lim_{r
\rightarrow r_{\rm th}^+}\, \pm\, \sqrt{e^{2\Lambda} - 1} = \, \infty.
\label{fo2} \end{equation}

\item As $l \rightarrow \pm \infty$
(or equivalently, $r \rightarrow \infty$), $e^{2\Lambda}
\rightarrow 1$ and $e^{2\Phi} \rightarrow 1$. This is the
asymptotic flatness condition on the wormhole spacetime.

\item  $\Phi(r)$ needs to be finite throughout the spacetime to ensure
the absence of event horizons and singularities.

\item  Finally, the  {\em flaring out} condition, that asserts that the
inverse of the embedding function $r(z)$ must satisfy $d^2 r/dz^2
> 0$ at or near the throat. Stated mathematically, \begin{equation}
-\frac{\Lambda^\prime\, e^{-2\Lambda}}{(1-e^{-2\Lambda})^2} > 0.
\label{fo1} \end{equation} This is equivalent to state that $r(l)$ has a
minimum.
\end{enumerate}

Static wormhole structures as those described by the above metric
require that the average null energy condition must be violated in
the wormhole throat. From the metric coefficients can be established (e.g. Morris and Thorne 1988, Visser 1996):



\begin{equation} G_{tt} + G_{rr} < 0,\end{equation}
where  $G_{tt}$ and $G_{rr}$ are the time and radial components of the Einstein tensor: $G_{\mu\nu}=R_{\mu\nu}-\frac{1}{2}g_{\mu\nu}R$.

This constraint can be cast in terms
of the stress-energy tensor of the matter threading the wormhole.
Using the field equations, it reads
\begin{equation}
\label{ha-c} T_{tt} + T_{rr} < 0, 
\end{equation} 
which represents a
violation of the null energy condition. This implies also a
violation of the weak energy condition (see Visser 1996 for
details). Plainly stated, it means that the matter threading the
wormhole must exert gravitational repulsion in order to stay
stable against collapse. Although there are known violations to
the energy conditions (e.g. the Casimir effect), it is far from clear at present whether large macroscopic amounts of ``exotic
matter'' exist in nature. If natural wormholes exist in the
universe (e.g. if the original topology after the Big-Bang was
multiply connected), then there should be observable
electromagnetic signatures (e.g. Torres et al. 1998). Currently, the observational data allow to
establish an upper bound on the total amount of exotic matter
under the form of wormholes of $\sim 10^{-36}$ g cm$^{-3}$. 
The production of this kind of matter in the
laboratory is completely out of the current technical
possibilities, at least in significant macroscopic quantities.

Finally, we mention that a wormhole can be immediately transformed
into a time machine inducing a time-shift between the two mouths.
This can be made through relativistic motion of the mouths (a
special relativity effect) or by exposing one of them to an
intense gravitational field (see Morris and Thorne 1988 and Morris et al. 1988 for further details; for the paradoxes of time travel, see Romero and Torres 2001).

\section{The future of black holes}

According to Eq. (\ref{age}), an isolated black hole with $M=10$ $M_{\odot}$ would have a lifetime of more than $10^{66}$ yr. This is 56 orders of magnitude longer than the age of the universe\footnote{We assume that the universe originated at the Big Bang, although, of course, this needs not to be necessarily the case.}. However, if the mass of the black hole is small, then it could evaporate within the Hubble time. A primordial black hole, created by extremely energetic collisions short after the Big Bang, should have a mass of at least $10^{15}$ g in order to exist today. Less massive black holes must have already evaporated. What happens when a black hole losses its mass so it cannot sustain an event horizon anymore?. As the black hole evaporates, its temperature raises. When it is cold, it radiates low energy photons. When the temperature increases, more and more energetic particles will be emitted. At some point gamma rays would be produced. If there is a population of primordial black holes, their radiation should contribute to the diffuse gamma-ray background. This background seems to be dominated by the contribution of unresolved Active Galactic Nuclei and current observations indicate that if there were primordial black holes their mass density should be less than $10^{-8}\;\Omega$, where $\Omega$ is the cosmological density parameter ($\sim 1$). After producing gamma rays, the mini black hole would produce leptons, quarks, and super-symmetric particles, if they exist. At the end the black hole would have a quantum size and the final remnant will depend on the details of how gravity behaves at Planck scales. The final product might be a stable, microscopic object with a mass close to the Planck mass. Such particles might contribute to the dark matter present in the Galaxy and in other galaxies and  clusters. The cross-section of black hole relics is extremely small: $10^{-66}$ cm$^{2}$ (Frolov and Novikov 1998), hence they would be basically non-interacting particles.

A different possibility, advocated by Hawking (1976), is that, as a result of the evaporation nothing is left behind: all the energy is radiated. This creates a puzzle about the fate of the information stored in the black hole: is it radiated away during the black hole lifetime or does it simply disappear from the universe?.

Independently of the problem of mini black hole relics, it is clear that the fate of stellar-mass and supermassive black holes is related to fate of the whole universe. In an ever expanding universe or in an accelerating universe as it seems to be our actual universe, the fate of the black holes will depend on the acceleration rate. The local physics of the black hole is related to the cosmic expansion through the cosmological scale factor $a(t)$, which is an arbitrary (positive) function of the co-moving time $t$. A Schwarzschild black hole embedded in a Friedmann-Robertson-Walker universe can be represented by a generalization of the McVittie metric (e.g. Gao et al. 2008):    

\begin{equation}
	ds^{2}=\frac{\left[1-\frac{2G M(t)}{a(t)c^{2}r}\right]^{2}}{\left[1+\frac{2G M(t)}{a(t)c^{2}r}\right]^{2}} c^{2}dt^{2}-a(t)^{2}\left[1+\frac{2G M(t)}{a(t)c^{2}r}\right]^{4} (dr^{2} +r^{2}d\Omega^{2}). \label{cosmicBH}
\end{equation}

Assuming that $M(t)=M_{0} a(t)$, with $M_0$ a constant, the above metric can be used to study the evolution of the black hole as the universe expands. It is usual to adopt an equation of state for the cosmic fluid given by $P=\omega\rho$, with $\omega$ constant. For $\omega<-1$ the universe accelerates its expansion in such a way that the scale factor diverges in a finite time. This time is known as the Big Rip. If $\omega=-1.5$, then the Big Rip will occur in 35 Gyr. The event horizon of the black hole and the cosmic apparent horizon will coincide for some time $t<t_{\rm Rip}$ and then the singularity inside the black hole would be visible to observers in the universe. Unfortunately for curious observers, Schwarzschild black holes surely do not exist in nature, since all astrophysical bodies have some angular momentum and is reasonable then to expect only Kerr black holes to exist in the universe. Equation (\ref{cosmicBH}) does not describe a cosmological embedded Kerr black hole. Although no detailed calculations exist for such a case, we can speculate that the observer would be allowed to have a look at the second horizon of the Kerr black hole before being ripped apart along with the rest of the cosmos. A rather dark view for the Doomsday.

\section{Conclusions}

Altogether, it is surely {\sl darker than you think}\footnote{Jack Williamson, Fantasy Press, 1948.}. 

\vspace{0.4cm}

\acknowledgments I would like to thank Ernesto Eiroa, Santiago E. Perez-Bergliaffa, and Brian Punsly for several discussions on black holes. My work on relativistic astrophysics is supported by the Argentine Agencies CONICET (PIP 5375) and ANPCyT (PICT 03-13291 BID 1728/OC-AR) and by the Ministerio de Educaci\'on y Ciencia (Spain) under grant AYA 2007-68034-C03-01, FEDER funds.

\end{document}